# Multi Weyl Points and the Sign Change of Their Topological Charges in Woodpile Photonic Crystals

Ming-Li Chang[1], Meng Xiao[1], Wen-Jie Chen[1], C. T. Chan[1]

*1 Department of Physics and Institute for Advanced Study, The Hong Kong University of Science and Technology, Hong Kong, China*

∗ Email: phchan@ust.hk

## Abstract

We show that Weyl points with topological charges 1 and 2 can be found in very simple chiral woodpile photonic crystals, which can be fabricated using current techniques down to the nano-scale. The sign of the topological charges can be tuned by changing the material parameters of the crystal, keeping the structure and the symmetry unchanged. The underlying physics can be understood using a tight binding model, which shows that the sign of the charge depends on the hopping range. Gapless surface states and their back-scattering immune properties are also demonstrated in these systems.

Recently, topological systems carrying Weyl points have attracted a lot of attention [1–15]. A Weyl point can be viewed as a magnetic monopole in momentum space that carries a topological charge and acts as a quantized source or sink of Berry flux [16]. As a nodal point, Weyl points can be viewed as the 3D counterparts of 2D Dirac points [17]. Dirac points can be gapped easily by breaking inversion or time reversal symmetry, while Weyl points are virtually indestructible, unless two Weyl nodes of opposite charges are tuned to annihilate each other. We know that PT broken symmetry systems can potentially carry Weyl points, but structural and materials details are important.

It is perhaps the lack of a simple rule to get Weyl points that leads to the structural complexity of classical wave systems designed to exhibit Weyl nodes [9,11] and it is unlikely that those structures can be made to realize Weyl point related phenomena at high frequencies. The natural question to ask is: Can we find simple periodic structures that carry Weyl points, preferably those structures that have been already made in the nanoscale. For high frequency applications, the most popular photonic crystal (PC) structures are the "woodpile" structures [18], composing of stacked layers of rods. Most of 3D PCs that have been fabricated to-date are indeed woodpile structures [19–21]. In this paper, we show the existence of Weyl points in very simple woodpile PCs which possess screw symmetry. Such chiral woodpiles structures can be fabricated [22] and are known to have interesting thermal properties [23]. We find that those simple structures not only support Weyl points with topological charge 1 but also carry higher order topological charge 2. In addition, topological charges distributions can be controlled by varying the permittivity distribution and topological charges can change sign even for structures with exactly the same symmetry. Such topological sign change can lead to the existence or absence of the gapless surface states on specific surface terminations.

The chiral woodpile PC has a unit cell as shown in Fig. 1(a), consisting of 3 layers of rods (cyan) with width $b$ and height $h$ and each is rotated 120° clockwise from the layer below, twisting up in the $z$ direction while forming a triangular-lattice in the $x$-$y$ plane. The length and height of a unit cell are $d$ and $p$, $p = 3h$. We first consider the air-in-metal configuration corresponding to air rods drilled into a perfect electric conductor (PEC) background, forming a connected network of waveguides with a specific screw symmetry. The computed eigenfield patterns show that fields of modes at low frequencies are confined to the crossing points in the rotation axis so that the band dispersion can be described reasonably well by the tight binding model (TBM) [24]. The minimal TBM that captures the connectivity of the network is shown in Fig. 1(b), showing three "atoms" stacked along the central rotation axis and connected by two kinds of nearest neighbor (NN) hopping: in-plane intra-layer hopping (blue bonds in Fig.1 (b)), and direct inter-layer hopping (orange bonds in Fig.1 (b)). We also include next nearest neighbor (NNN) hopping and an example of such hopping is marked by a red bond in Fig.1 (b). The TBM Hamiltonian of the system is

$$H = \sum_{i,k} \varepsilon a^\dagger_{i,k} a_{i,k} + \sum_{\langle i,j \rangle} t_{n1} a^\dagger_{i,k} a_{j,k} + h.c. + \sum_{i,k} t_{n2} a^\dagger_{i,k} a_{i,k+1} + h.c. + \sum_{\langle\langle i,j \rangle\rangle} t_{nn} a^\dagger_{i,k} a_{j,k+1} + h.c. \quad (1)$$

where $a$ ($a^\dagger$) are the annihilation (creation) operators, and the first and second subscript represent the position in each layer and the layer number respectively. The subscript $\langle i,j \rangle$ represents intra-layer NN lattice, and $\langle\langle i,j \rangle\rangle$ denotes the NNN lattice as shown in Fig.1 (b). The first term of the Hamiltonian represents the on-site energy while the second terms are the NN intralayer hopping with hopping strength $t_{n1}$ and the third terms are NN interlayer hopping with hopping strength $t_{n2}$. The last terms represent the NNN hopping. The corresponding Hamiltonian in the momentum space can be written as:

$$H(\mathbf{k}) = \begin{pmatrix} t_{n1}\alpha & e^{-ik_zh}(t_{n2}+t_{nn}(\alpha+\beta)) & e^{ik_zh}(t_{n2}+t_{nn}(\alpha+\gamma)) \\ e^{ik_zh}(t_{n2}+t_{nn}(\beta+\alpha)) & t_{n1}\beta & e^{-ik_zh}(t_{n2}+t_{nn}(\beta+\gamma)) \\ e^{-ik_zh}(t_{n2}+t_{nn}(\gamma+a)) & e^{ik_zh}(t_{n2}+t_{nn}(\gamma+\beta)) & t_{n1}\gamma \end{pmatrix}, \quad (2)$$

where $\alpha = 2\cos(k_y d)$, $\beta = 2\cos((\sqrt{3}k_x - k_y)d/2)$, $\gamma = 2\cos((\sqrt{3}k_x + k_y)d/2)$.

$\Gamma A$ and $KH$ are the directions with screw rotational symmetry, which ensures that two bands are degenerate at one frequency at $K$ and $\Gamma$ points (See the Supplemental Material Sec. 2). To investigate the topological properties of the bands, we apply the **k·p** method around the $K$ and $\Gamma$ points. As a third band locates at a different frequency, the $3\times3$ Hamiltonian can be reduced to a $2\times2$ Hamiltonian to describe the dispersions near the degenerate point (See Supplementary material Sec.1) [25] as:

$$H(\mathbf{K}+\mathbf{p}) = (2t_{nn} - t_{n1} - t_{n2})\sigma_0 + (t_{n1} - 2t_{nn})d(p_x - 3p_y)\sigma_1/4 \\ + \sqrt{3}(t_{n1} - 2t_{nn})hp_z\sigma_2 - \sqrt{3}(t_{n1} - 2t_{nn})d(p_x + p_y)\sigma_3/4 \quad (3.a)$$

$$H(\Gamma+\mathbf{p}) = \left(2t_{nn} - t_{n1} - 4t_{nn} - (t_{n1} - 2t_{nn})d^2(p_x^2 + p_y^2)/2\right)\sigma_0 \\ + (t_{n1} - 2t_{nn})d^2\left(\sqrt{3}p_x^2 - 2p_xp_y - \sqrt{3}p_y^2\right)\sigma_1/8 \quad (3.b) \\ + \sqrt{3}(t_{n2} + 4t_{nn})hp_z\sigma_2 + (t_{n1} - 2t_{nn})d^2\left(-p_x^2 - 2\sqrt{3}p_xp_y + p_y^2\right)\sigma_3/8$$

where $\sigma_1$, $\sigma_2$, $\sigma_3$ are the Pauli's matrixes, $\sigma_0$ is a $2\times2$ identity matrix and $p_x$, $p_y$, $p_z$ are components of **p** which denotes the momentum near the degeneracy point. The reduced Hamiltonian is in the Weyl form with linear terms of $p_x$, $p_y$ and $p_z$ at $K$ while second order terms of $p_x$, $p_y$ and linear term of $p_z$ at $\Gamma$. Similar results can be obtained at $H$ and $A$ points (See the Supplemental Material Sec. 1). These TBM results indicate two Weyl points with topological charge $\pm 1$ can be found at $K$ and $H$ and two Weyl points with topological charge $\pm 2$ are at $\Gamma$ and $A$. [26] The multi Weyl points are protected by screw symmetry and the topological charge can be calculated from the Chern number over a closed surface around these points [27]. The sign of charges at the high symmetry points are

$$C(K) = \text{sgn}(t_{n2} - 2t_{nn}), \quad (4.a)$$

$$C(\Gamma) = -2\text{sgn}(t_{n2} + 4t_{nn}). \quad (4.b)$$

While they take the opposite values at $H$ and $A$, i.e.,

$$c(H) = -\text{sgn}(t_{n2} - 2t_{nn}), \quad (4.c)$$

$$c(A) = 2\text{sgn}(t_{n2} + 4t_{nn}). \quad (4.b)$$

(See Supplemental Material Sec. 1). These TBM results indicate that Weyl points of topological charge of 1 and 2 can be found respectively at the zone boundary and the zone center, and the sign of the topological charge can be tuned by controlling $t_{n2}$ and $t_{nn}$.

To show that Weyl points can indeed exist in the real optical system, we calculate the photonic bands of the chiral woodpile air-in-metal PCs with $b = 0.4d$ and $h = 0.5d$ using COMSOL as shown in Fig. 2 (a)-(e). The lower two bands form a single Weyl point at $K$ and a double Weyl point at $\Gamma$ at $k_z = 0$, while Weyl points of opposite topological charges are formed by the upper two bands at $H$ and $A$ at $k_z = \pi / p$. The 3D band structures are shown in Fig. S1 of the Supplemental material. When $k_z \neq n\pi / p$, the degeneracies are lift along the $k_z$ axis as shown in Fig. 2 (d) and (e). The current system can be well-described with the TBM as shown in Fig. S2 (a), where we compare the band structures of the whole Brillouin zone obtained with COMSOL and TBM using fitting parameters $\varepsilon = 0.804c/d$, $t_{n1} = -0.011c/d$ and $t_{n2} = 0.0675c/d$ and the NNN hopping terms are ignored. Furthermore, the phase of screw rotation eigenvalues of three interested bands evolves from $-\pi$ to $\pi$ as shown in Fig.S2 (b) and (c) by the color code. Different representations of screw rotation operators enable the band crossings at the Brillouin zone boundaries and the eigenvalues of screw symmetry of different bands connect to each other at the $k_z$ zone boundary [28].

The topological charges, as sources or sinks of Berry flux, can lead to gapless boundary modes, whose existence can be determined by analyzing the topological charge distributions. For the air-in-metal configuration, the effective NNN hopping $t_{nn}$ is essentially zero (wave cannot penetrate metal), and the sign of topological charges at the high symmetry points depends only on $t_{n2}$. The topological charge distributions obtained numerically for the real PC are shown in Fig.2 (f). We note that TBM (Eq. 4) gives the same distribution. At the $k_z = 0$ plane (Fig. 2(a)), Weyl nodes with charge +1 are found at *K* and *K'* (yellow dot), while Γ has a node with charge -2 (blue dot). At the $k_z = \pm\pi/p$ plane, as shown in Fig. 2(c), Weyl nodes with charge -1 are located at *H* and *H'* (cyan dots), and *A* has a Weyl node with charge +2 (red dot). We note that the total charge at the $k_z = 0$ or $k_z = \pm\pi/p$ planes is zero and if we cut any planes perpendicular to the $k_z$ axis, the total Berry flux passing through these planes vanishes. The Chern number is hence zero for all the 2D bands with a fixed $k_z$ and topological gapless surface states cannot be found on truncated boundaries parallel to the *z*-direction.

The Weyl nodes in this chiral woodpile structure are protected by the $\tilde{C}_3$ screw symmetry. For the exactly the same symmetry, the distribution and the charge of the Weyl node actually depends on the material parameters. We will now show that by changing the air-in-metal configuration to the inverse configuration of metal-in-air, while keeping the geometry the same, some Weyl nodes changes sign with the consequence that the total Berry flux going through any band with fixed $k_z$ no longer vanishes. From viewpoint of a TBM, Eqs. (4.a) and (4.b) show that when the NNN hopping is comparable to the inter-layer hopping, sign change of topological charges at *K* or $\Gamma$ can occur. Referring to Fig. 1(a), the NNN hopping is essentially zero if the blue rods are air and the background is metal, but for the inverse structures in which the blues rods are metal and background is air, wave can travel in the background meaning that the NNN hopping cannot be zero. As such, the Weyl nodes charge can potentially change sign and of course whether the sign change can happen needs to be verified by a full wave calculation for the real system, which is shown in Fig. 3. The band structures of the reduced Brillouin zone at $k_z = 0, 0.5\pi/p, \pi/p$ are shown in Figs. 3 (a)-3(c) for the metal-in-air configuration with the width and height of metal rods being $b = h = 7.2 \mu m$ and the lattice constant $d = 60 \mu m$. A single Weyl point exists at 3.5THz (yellow dot in Fig. 3(a)) formed by the intersection of red and blue bands at *K* while the opposite charge nodes exists at *H* (cyan dot in Fig. 3(c)). We find two double Weyl points with opposite topological charges at $\Gamma$ (red dots in Fig. 3(a)) and *A* (blue dots in Fig. 3(c)) respectively with two bands carrying quadratic dispersions in the $k_x - k_y$ plane. When $k_z = 0.5\pi/p$, the degeneracies are lifted and two gaps are formed as shown in Fig. 3(b). The topological charge distribution is shown in Fig. 3 (d). Compared with Fig. 2(f) (for air-in-metal), we see that the charge 2 topological nodes have changed sign, or equivalently, one may say that the distribution of topological changes has changed. The topological properties of the bands and gaps can be obtained by analyzing the distributions of topological charges. We note that a positive topological charge acts as the source of Berry flux for the upper band and a sink for the lower band and vice versa. Fig. 3(e) shows the source/sink distributions for the red band (the lower band), which shows that charge 1 flux sinks locate at the $k_z = 0$ plane and topological charges 2 flux sources locate at $k_z = \pm \pi / p$ planes, generating Berry flux between these planes. Fig. 3(f) and Fig. 3(g) shows respectively the source/sink distribution for the green band (the upper band) and the blue (middle) band, showing net Berry flux going through $k_z = constant$ planes for the green band while no net flux passes through $k_z = constant$ planes for the blue bands. As a consequence, the Chern number of the red and green band are -1 and +1 while it is 0 for the blue (middle) band when $k_z > 0$, which in turn implies that the gaps in Fig.3 (b) are topological nontrivial and gapless boundary modes should exist in these gaps. To investigate the gapless surface states, we cut two surfaces perpendicular to *x* axis, resulting in a strip (see Fig.4 (a)) that is periodic along the *y* and *z* directions. We cap the truncated surface by PEC boundaries, which serve as a topological trivial gap material. For $k_z = 0.5\pi / p$, we find two gapless surface states in each of these two topological non-trivial gaps as shown in Figs. 4(b) and 4(c). The norm of *E* field patterns of two surface states show that the fields for the blue band are localized at the right boundary, propagating along +*y* (see Figs. 4(d) and 4(f)), while the fields for the red band are localized at the left boundary, propagating along -*y* (see Figs. 4(e) and 4(g)).

The surface states should be robust against backscattering when $k_z$ is preserved. To demonstrate this explicitly, we construct a zigzag boundary between two woodpile PCs as shown in Fig.5. The upper PC has the air-in-metal while in the lower PC has the metal-in-air configuration. The width of air rods and metallic rods are $33\mu m$ and $12\mu m$ respectively. The height of both are $9\mu m$. The in-plane lattice constants are the same for both parts, which is $d = 60nm$. According to our analysis above, these two woodpile structures should process different topological gap properties. In the simulation, the top, left and right boundaries are set as PEC boundary and the bottom boundary is set as scattering boundary so that waves can lead outside from the bottom boundary. The sample is periodic along the *z* direction. In Fig. 5(a), we place a line source at the left hand side as marked by the red star and set $k_z = -0.5\pi/p$. We set $f = 3.5THz$, which is inside the common gap for both woodpile structures. The excited surface state first propagates along the +*y* direction on the left boundary, and then turns around the corner when it meets the metal-in-air woodpile and goes along the zigzag boundary from left to right. No backscattering occur even the boundary bends dramatically and the edge state just leaks outside from the bottom boundary. We move the source to the right hand side as shown by the red star in Fig.5 (b) and set $k_z = 0.5\pi/p$. At the same frequency, the wave now propagates along the zigzag boundary from right to left without backscattering. We note that if we use Ag instead of PEC, the simulation shows almost the same results at $f = 3.5THz$.

In summary, we show that the band structure of metallo-dielectric chiral woodpile PCs exhibit Weyl nodes with topological charges of 1 and 2. Gapless surface states supporting robust transport are found at the surface of these woodpile PC slabs. The physics can be understood using an analytic TBM, which shows that the sign of the topological charge can be changed by controlling the range of the hopping parameters. This interesting effect is verified in numerical calculations. We note that pure Si woodpile structures also carry Weyl nodes, as shown in Fig. S4 and Fig. S5 (See Supplemental Material Sec. 3). Woodpile structures have been fabricated down to the nano-scale, making them promising platforms for achieving Weyl point related phenomena at the high frequency regimes.

We thank Wen-yu He for discussion. This work was supported by Research Grants Council, University Grants Committee, Hong Kong (AoE/P-02/12).

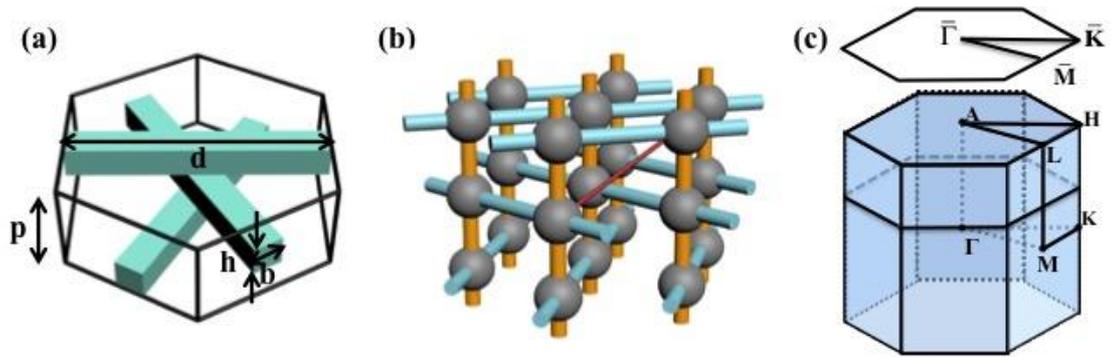

Fig.1 (Color Online) (a) The unit cell contains three layers of rods twisting up along the z direction. (b) The tight binding model with "atoms" connected by "bonds". Blue and orange bonds represent respectively the intralayer and interlayer nearest neighbor coupling. The red bond marks a next nearest neighbor hopping. (c) The reciprocal space of (a).

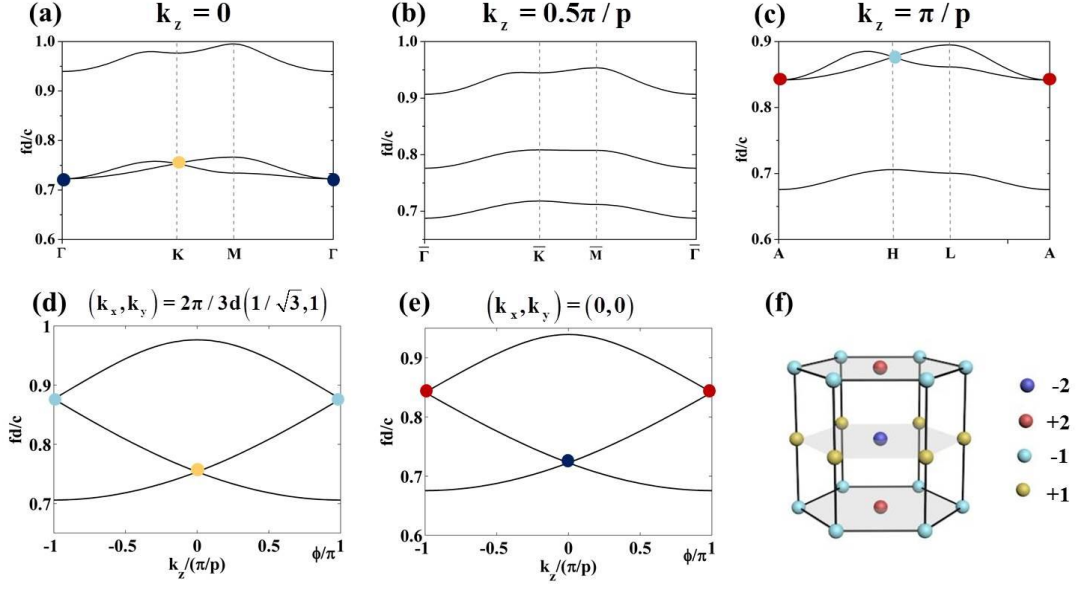

Fig.2 (Color Online) The band structures of air-in-metal crystals calculated by COMSOL for (a) $k_z = 0$; (b) $k_z = 0.5\pi/p$; (c) $k_z = \pi/p$. Weyl points are found at $K$, $H$, $\Gamma$ and $A$. (d) and (e) show the band structures along $KH$ and $\Gamma A$, respectively. The corresponding 3D band structures are shown in Fig. S1 of the appendix. (f) The topological charges distributions, showing the $\pm 2$ charges at $\Gamma$ and $A$, and $\pm 1$ charges at $K$ and $H$.

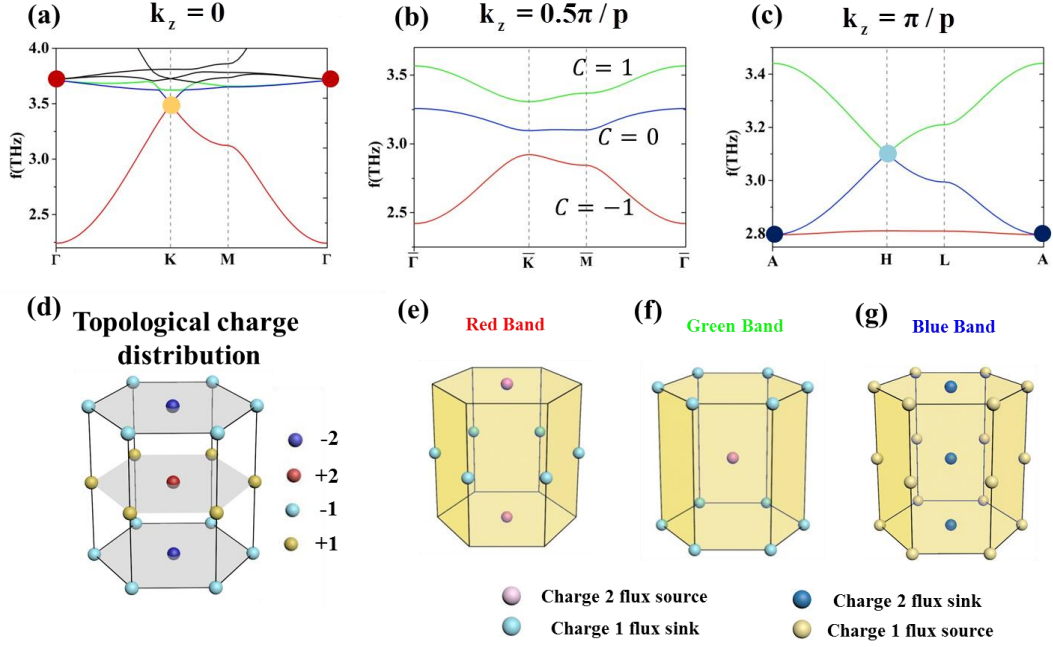

Fig.3 (Color Online) The band structures of the metal-in-air configuration for (a) $k_z = 0$; (b) $k_z = 0.5\pi/p$; (c) $k_z = \pi/p$. Single Weyl points are found at *K*, *H* and double Weyl points are found at $\Gamma$, *A*. In (b), the bands are marked with the calculated Chern numbers. (d) Topological charge distributions showing that the net charges at the plane $k_z = n\pi/p$ do not vanish. Weyl points, as sources/sinks of Berry flux, are plotted in (e), (f) and (g). (e) For the red band, the Weyl points with -2 charge is a flux source at *A* and the +1 charge is flux sink at *K*. No zero Berry flux goes through the planes between $k_z = 0$ and $k_z = \pm\pi/p$. Similar effect can be found for the green band, shown in (f). This gives Chern numbers of -1/+1 for the red/green bands. (g) For the blue band, the charge 1 Weyl point at *K* and *H* acts as a flux source and the charge 2 Weyl point at $\Gamma$ acts as a flux sink at the $k_z = n\pi/p$ planes, resulting in no net in flux at any $k_z$ planes and zero Chern number for the blue band.

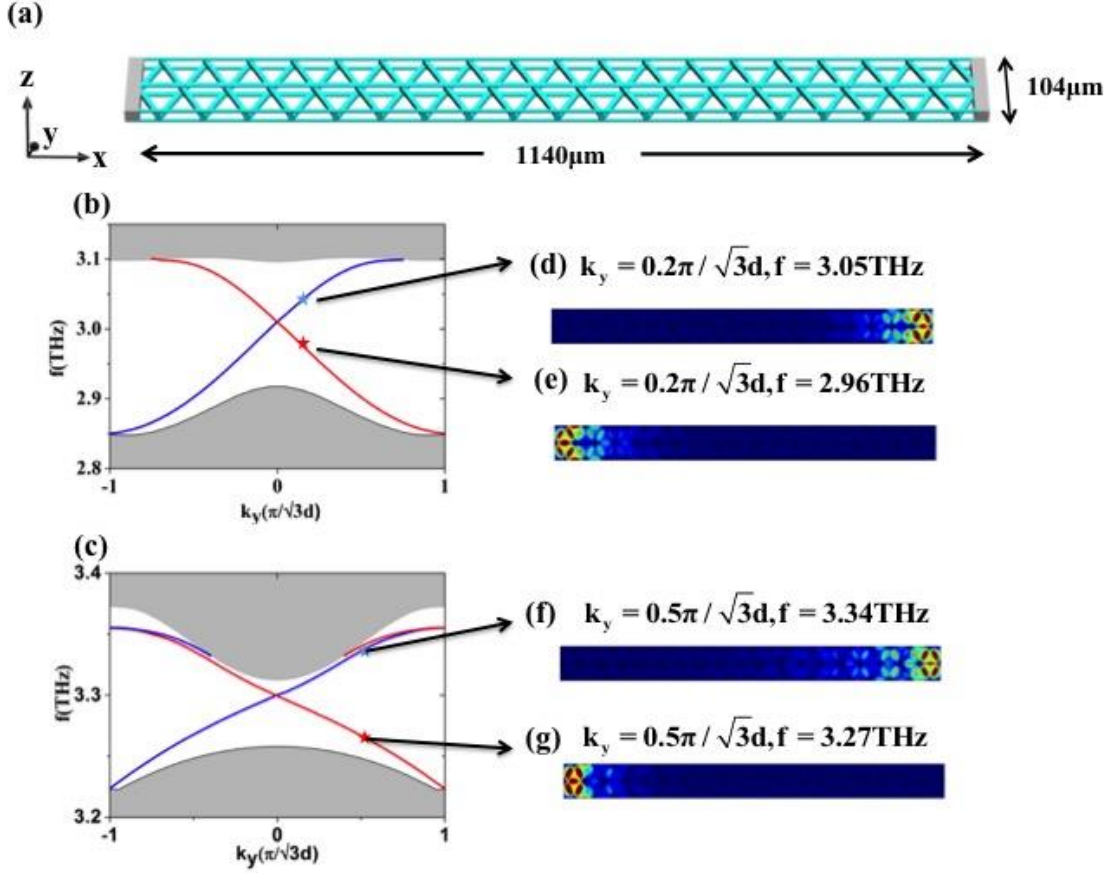

Fig.4 (Color Online) (a) The slab is periodic along *y* and *z* and sandwiched by PEC slabs in the *x* direction. The length and width of the slab are $1140nm$ and $104nm$. (b) The projected band structure, where $k_z = 0.5\pi/p$ and the parameters used here are the same as those in Fig. 3. Gapless surface states (blue and red bands) exist in the gap around 3THz. The bulk bands are represented by grey regions. Panels (d) and (e) show the eigen $|E|$ fields of the boundary mode marked by the blue and red stars respectively at $k_y = 0.2\pi/\sqrt{3}d$. The positive group velocity mode (blue) is localized on the right surface, while the negative group velocity mode (red) is localized on the left. (c) Another two surface states exist in the higher gap around 3.3THz. (f), (g) show $|E|$ of the surface states marked by the blue and red stars in (c) with $k_y = 0.5\pi/\sqrt{3}d$.

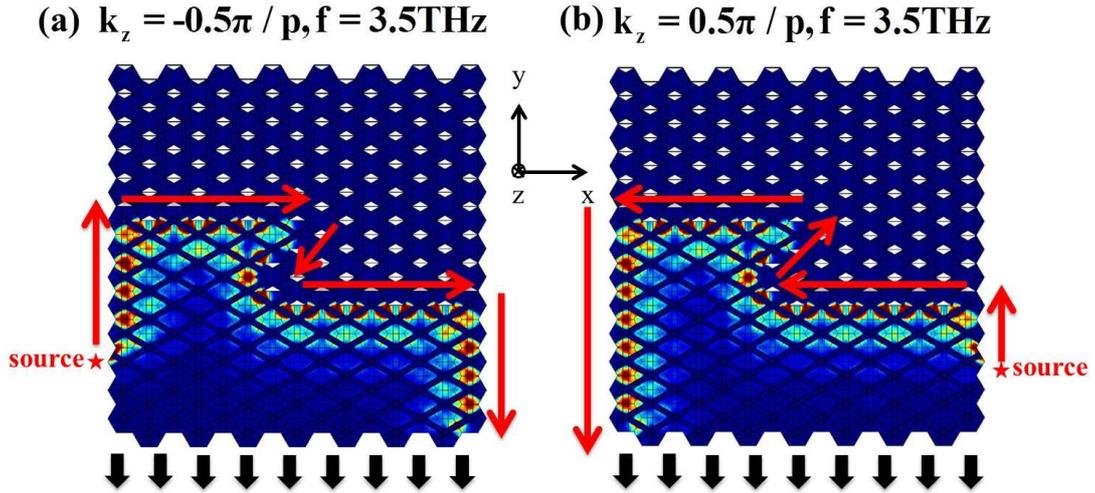

Fig.5 (Color Online) One-way transport along the boundary between two woodpile structures with different topological properties. The structure above the zig-zag boundary has the air-in-metal configuration while that below the boundary has the metal-in-air configuration. (a) When $k_z = -0.5\pi/p$, the edge state propagates from left to right; (b) When $k_z = 0.5\pi/p$, the edge state propagate from right to left. The positions of sources are indicated by the red stars. No backscattering can happen in both cases.